\documentclass[twocolumn,pra,nofootinbib,superscriptaddress,showpacs,10pt,aps,draft]{revtex4-1}

\usepackage{color}

\usepackage{amsmath}
\usepackage{amssymb}
\usepackage{graphicx}
\usepackage{color}

\usepackage[T1]{fontenc}
\usepackage[utf8]{inputenc}

\usepackage{amsthm}
\usepackage{bbm}

\newcommand{\tr}[1]{\mathrm{tr}\left[{#1}\right]}

\def\idty{{\leavevmode\rm 1\mkern -5.4mu I}} 

\def\Rl{{\mathbb R}}

\let\eps\varepsilon

\def\braket#1#2{\langle #1,#2\rangle}

\def\tr{\mathop{\rm tr}\nolimits}

\begin{document}
\title{Measurement Uncertainty: Reply to Critics}

\author{Paul Busch}
\email{paul.busch@york.ac.uk}
\affiliation{Department of Mathematics, University of York, York, United Kingdom}

\author{Pekka Lahti}
\email{pekka.lahti@utu.fi}
\affiliation{Turku Centre for Quantum Physics, Department of Physics and Astronomy, University of Turku, FI-20014 Turku, Finland}

\author{Reinhard F. Werner}
\email{reinhard.werner@itp.uni-hannover.de}
\affiliation{Institut f\"ur Theoretische Physik, Leibniz Universit\"at, Hannover, Germany}

\date{\today}
\begin{abstract}
In a recent publication [PRL 111, 160405 (2013)] we proved a version of Heisenberg's error-disturbance tradeoff.
This result was in apparent contradiction to claims by  Ozawa of having refuted these ideas of Heisenberg.
In a direct reaction [arXiv:1308.3540] Ozawa has called our work groundless, and has claimed to have found both a counterexample and an error in our proof.
Here we answer to these allegations. We also comment on the submission [arXiv:1307.3604] by Rozema {\it et al}, in which our approach is unfavourably compared to that of Ozawa.
\end{abstract}
\maketitle

\section{Introduction}
Uncertainty relations are a key insight of quantum theory, going back to Heisenberg's seminal paper \cite{Heisenberg1927} from 1927. It therefore must come as a surprise to many that over 85 years later there can still be a controversy about them. In fact, there is no controversy about the basic textbook version (``preparation uncertainty''), which forbids quantum states that have too sharp distributions for both position and momentum. Ironically, this aspect is not even discussed in Heisenberg's paper, who instead considers in his famous microscope example the trade-off between the accuracy of an approximate position measurement and the disturbance of momentum which the measurement causes. This can be generalized to a statement about the impossibility of accurate joint measurements of two complementary observables (``measurement uncertainty''). It is this aspect which is controversial, and while Ozawa \cite{Ozawa03,Ozawa04,Ozawa05} claims to have refuted Heisenberg's ideas we \cite{Werner04,BLW2013a,BLW2013b,BLW2013c} have proved a general result confirming his intuition.

This apparent contradiction is not a mathematical one, since the two approaches start from different definitions of ``error'' and ``disturbance''.
Ozawa's approach \cite{Ozawa03,Ozawa04,Ozawa05}, which grew out of the work of Arthurs 1965 \cite{AK65}, is based on the expectations of squared differences of certain operators (``noise operators'') in a measurement scheme. These quantities depend on the input state of the measurement. Ozawa claims that a certain inequality written in such terms (\eqref{false} below) is ``Heisenberg's inequality'', and shows that it is wrong in general. The definitions in our approach \cite{BLW2013c} 
are characteristic of the measurement scheme, and hence independent of the input state. We prove for these quantities a relation that looks like the standard textbook form of uncertainty relations, even though the $\Delta$-quantities have quite a different meaning. We find that this identifies one aspect of Heisenberg's paper, which has a general, rigorous quantitative interpretation. In contrast, far from ``disproving Heisenberg'', Ozawa has simply failed to identify such a grain of truth.

Ozawa has written an attack on our approach \cite{Ozawa2013}. This was worded so aggressively, and was so low on scientific quality, that it would seem to be more of an embarrassment to the author than an argument requiring an answer. Nevertheless, as mathematical physicists we have to respond, since he claimed to have found a counterexample to our theorem and also claimed to have spotted an error in our proof. Such a reply is the main content of this note. We also reply to a critique \cite{Roz13} by one of the experimental groups backing Ozawa.

We have laid out our criticism of the Ozawa approach and, more generally of ``noise operator'' definitions of error and disturbance in a separate paper \cite{BLW2013a}, which we very briefly summarize and in some ways extend in the next section. For our Theorem with a sketch of proof we refer to  \cite{BLW2013c}, for a full proof to \cite{BLW2013b}, and for a discussion of the qubit case, which is the basis of the experiments, to \cite{BLW2014}.

\section{Points of a debate}
It is the normal state of affairs that intuitive concepts can be sharpened mathematically in several different ways. Also notoriously non-unique are quantum analogs of classical concepts. All one can do in such a case is to state the options as clearly as possible, and point out contexts in which one may be more appropriate than another. Basically, measurement uncertainty is just another example of this. If you want to discuss operationally defined figures of merit for measurement devices like the ``resolution of a microscope'', we recommend our definitions and results. If you hope to get more detailed statements depending on the input states, for example for purposes of cryptography, you may consider entropic ideas \cite{Furrer}, and Ozawa's might seem a step in the direction of ``more detailed'' statements. However, there are some requirements which cannot be put aside easily as matters of taste or choice of an appropriate context.
Before coming to these, we begin by sketching the context: Ozawa's and our definitions of the terms in the respective uncertainty relations.

\subsection{The basic definitions}
Ozawa considers a model of measurement by which the system is coupled by a unitary operator $U$ to a probe in initial state $\sigma$, which is also a canonical degree of freedom. The position outcome is identified with the position of the probe after the measurement $U^*(\idty\otimes Q)U$. Similarly the momentum of the system before and after the measurement are represented by the operators $P\otimes\idty$ and $U^*(P\otimes\idty)U$, respectively. The Ozawa error $\epsilon_O$ and the Ozawa disturbance $\eta_O$ are then defined by
\begin{eqnarray}\label{errO}
    \epsilon_O^2&=& \tr\rho\otimes\sigma \left(U^*(\idty\otimes Q)U-Q\otimes\idty\right)^2 \\
        \eta_O^2&=& \tr\rho\otimes\sigma \left(U^*(P\otimes\idty)U-P\otimes\idty\right)^2. \label{distO}
\end{eqnarray}
Then in general the ``uncertainty relation''
\begin{equation}\label{false}
  \epsilon_O \eta_O\geq \hbar/2 \end{equation}
is false.

We consider a general joint measurement device, i.e., any device which outputs both a momentum value and a position value, which in the above setting would be the position of the probe and the momentum of the object particle after the measurement interaction. We say that the position accuracy of such a device is $\delta_Q$, if for arbitrary input state $\rho$ with sharp position at $q$, the distribution of the position outputs has a quadratic deviation at most $\delta_Q^2$ from the ``correct'' value $q$. Then, for any measuring device $\delta_Q\delta_P\geq\hbar/2$, and equality is attained for a device whose output distribution is the Husimi phase space distribution.

\subsection{Do Ozawa's definitions capture Error and Disturbance?}
If one defines a quantity like ``error'' of a measurement, then the mathematical concept should share the basic features of the intuitive concept. This is usually tested most clearly on extreme cases: When a device gives an output distribution different from that of an ideal reference, it is clearly in error, and the error quantity defined formally should not assign error zero to this situation. Conversely, a device for which input state and output state are the same can hardly be claimed to have disturbed the system, and so we should expect the definition to assign disturbance zero. Neither criterion is met by Ozawa's definitions \cite{BLW2013a}. If anything they express some combination of variances of the device and of the input state. So the criticism in this case is that the words ``error'' and ``disturbance'' are misnomers.

\subsection{A failed quantum generalization}
In the classical case, two random variables $f,g$ for which $(f-g)^2$ has expectation zero coincide with probability one, and hence they have the same distribution. In the quantum case this fails in general, i.e., if for some state $\Psi$ we have $\braket\Psi{(A-B)^2\Psi}=0$ we have $A\Psi=B\Psi$, but when $A$ and $B$ do not commute on the state $\Psi$,
the higher moments, and hence the distributions of $A$ and $B$ will not, in general, coincide.
This is the reason for some of the failures mentioned in the previous paragraph. Ozawa likes to cite Gau\ss\ for justifying the ``expected squared difference'' expressions \eqref{errO} and \eqref{distO}, completely ignoring the non-commutativity. This is a inexcusable oversight, especially in a discussion of quantum uncertainty.

By the rules of the book, measuring $(A-B)^2$ requires setting up a measurement for which the probability for outcomes in an interval is given by the expectation of a spectral projection of $(A-B)$. There is no direct way implementing this given $A$ and $B$, and in any case such a measurement is not compatible with either $A$ or $B$ (unless these observables commute). In other words, the Ozawa error of a microscope refers to a measurement incompatible with the operation of the microscope itself. This is what we mean by saying that his definitions are {\it not operational}: they do {\em not} reliably indicate the presence, absence, or the extent of differences between the distributions of the target observable and the intended approximating observable. It is still possible to design a measurement for the expectation of $(A-B)^2$, as for any operator, but this does not cure the basic flaw in the definition.

Presumably in response to our non-operationality criticisms in \cite{Werner04} and \cite{BuHeLa04},
Ozawa \cite{Ozawa04} offered the so-called 3-state method for determining his error quantity from the statistics of the POVM being measured as an approximate measurement of a given target observable.
However, this approach undermines the intended interpretation of Ozawa's error quantity as being state-specific: the very fact that three distinct states are to be measured in order to obtain its value takes away any justification for associating this quantity with any particular state. This failure of Ozawa's definition becomes strikingly evident in the qubit case discussed at the end of this note.

\subsection{State dependence}
The left hand side of \eqref{false} is not bounded below by a positive constant. In fact it is easy to come up with examples once one realizes that this failure is typical for state-dependent measures \cite{Rud2013}. Basically, one can take a standard destructive nearly ideal position measurement followed by the re-preparation of the input state. In this way both error and disturbance can be made arbitrarily small. The disturbance becomes evident on other input states, which is why we consider a supremum over input states at this point. A non-disturbing measurement to us is one which leaves {\it every} state unchanged, not one that leaves {\it some} state unchanged. So the failure of the error-disturbance product to be bounded below is just due to an unfortunate choice of quantifiers, quite apart from the other flaws of the definitions. Nothing can be learned from this failure about quantum physics.

This has been noted by Appleby \cite{Appleby1998a,Appleby1998b} in 1998 for the noise operator approach, who took this as a motivation to move on to a state-independent approach with worst case errors, in this regard much like ours. This was some years before Ozawa started his ``refuting Heisenberg'' programme. Hence it can be said that in this respect Ozawa's contribution was a step backwards on Appleby's insight. To be sure he produced some correction terms which turn \eqref{false} into a correct inequality, for which he uses the odd\footnote{For mathematical physicists the attribute ``universally valid'' normally goes without saying.}
phrase ``universally valid'' inequality.

The false inequality \eqref{false} plays a key role in Ozawa's recent writings, namely as ``Heisenberg's'' error-disturbance relation. In this way Appleby's simple observation that it is false becomes Ozawa's ``disproof'' of Heisenberg's paper \cite{Heisenberg1927}, and the experimental verification of this simple fact creates the impression that Heisenberg is in conflict with Nature herself. This created a considerable media-hype in 2012. In order to get a balanced view of this we clearly have to take a look at Ozawa's justification of blaming Heisenberg for \eqref{false}.

\subsection{Is it Heisenberg's?}
So what about historical accuracy? We are perfectly aware that our definitions of error and disturbance are not in Heisenberg's writings anywhere. Hence our relations are not Heisenberg's either. They are closer to Heisenberg's paper than the usual textbook version, since he clearly discusses measurement uncertainty rather than preparation uncertainty. In order to emphasize this, and also to honour the deep intuitions in his paper, we have, on occasion called our relations ``Heisenberg relations''. This is covered by the usual naming practices in science.

Ozawa's case is different, however. His claim of having refuted Heisenberg rests completely on putting words -- his own false inequality \eqref{false} -- into Heisenberg's mouth. This is absolutely ``groundless''. Heisenberg does not even write his relations with an inequality sign but with a mathematically unexplained tilde. He is purposefully vague, in keeping with the program outlined in the introduction of his paper, of giving an intuitive, heuristic understanding of why orbits must be criticised as unobservable in an atom (his discovery, just 2 years old then), but why we can still see tracks in a cloud chamber. This intuition stands as a valid insight and, as we have shown, one can even back it up by a precise quantitative statement. Heisenberg was probably not too interested in rigorous formulations. Although he announces a proof to be given later in the paper (on the basis of ``his'' commutation relations), he famously never gave one, and probably his grasp of the formalism in 1927 would not have been up to it.

Ozawa's justification for laying his mistake on Heisenberg is that Heisenberg uses the phrase ``mean error'' at some point, which in the context hardly means more than the admission that this is not a sharply defined quantity. There is no indication that Heisenberg was thinking in terms of a specific average, or any operator difference for the error. In fact, for modern readers it is clear that he is still quite far from  taking abstract operators as basic entities of quantum mechanics, which he considers only through the veil of Dirac's transformation theory.

Another argument given by Ozawa is his claim that the root-mean-square measurement error (``noise'') and  disturbance are ``naturally'' or ``standardly'' defined by his $\epsilon_O,\eta_O$ \cite{Ozawa05}. In his ``disproving'' manuscript \cite{Ozawa2013} he implies that his definition is the only possible quantum generalization of Gau\ss's classic concept of root-mean-square error. This is reiterated in the supplement of \cite{Ozawa-etal2014}. Again Ozawa describes the classical case, and Gau\ss's remarks on the topic. He shows that formula \eqref{errO} holds in the case of commuting observables. Then comes a characteristic leap in the argument: The same formula is applied, without even so much as a comment, to the non-commutative case. So even if one is willing to accept Gau\ss's definition as part of the Canon, the non-commutative extension is entirely Ozawa's invention, not justified by him,  and, as we have shown, indeed highly problematic. And even if one might hypothesize that Heisenberg, had he cared about making his notion precise, would have tried Gau\ss's canonical notion, the claim that he wouldn't have seen or would have chosen to ignore the typical problems brought in by quantum mechanics, and would thus have translated the ``$q_1$'' of his paper to Ozawa's \eqref{errO} is, at best, an argument by  lack of imagination.

Ozawa's advertising and that of the associated experiments largely rests on the contrast between the false relation \eqref{false} consistently referred to as Heisenberg's on the one hand and the corrected one on the other. This is in stark contrast to the lack of historical argument for this attribution. We have tried \cite{BLW2013a} to explain some of the richness in the intuitive content of \cite{Heisenberg1927}. But if later papers \cite{Vienna2,Baek13,Ozawa-etal2014} are any indication we are afraid that Ozawa is not interested, and continues to try to push the point by mechanical  repetition. So in contrast to all the argument-free rhetoric we have to insist that the only person ever to advance the false relation \eqref{false} for serious consideration is Ozawa. As an aid to understanding what his papers are about we suggest to any potential reader to replace the name Heisenberg by Ozawa, whenever it comes up in relation to \eqref{false}. This leads to a rather more adequate representation of the actual scientific content. 

Overzealous journalists have taken the false advertising even further by claiming that this ``disproof'' of Heisenberg's uncertainty would somehow render quantum cryptography insecure. Indeed, uncertainty relations play a key role in some modern security proofs \cite{Tomamichel2012,Berta-etal2010,Furrer}, but the statements used there are proven theorems. How could an error committed elsewhere (by Heisenberg, Ozawa, or whoever) have any bearing on such results?

In any case, despite the severe differences between Ozawa's and our take on Heisenberg, there is a common core: both parties seem to agree that the logical structure of Heisenberg's measurement uncertainty principle is that of a mutual trade-off between measurement accuracies (or accuracy vs.\ disturbance) that is dictated by the incompatibility of the observables under investigation. The focus on  disproving Heisenberg surely has helped to bring this fundamental principle once more into the public limelight. However,  the scientific question at stake is that of an appropriate, {\em operationally significant} and  mathematically rigorous, formulation of the principle {\em as a consequence of quantum theory}.

\subsection{Do experiments help?}
An interesting point of the recent debate is that the failure of Ozawa's relation \eqref{false} has been verified experimentally. This was touted as a confirmation of his theory, a refutation of Heisenberg by Nature herself, and contributed greatly to the hype. But actually an experiment does nothing to ``cure'' a bad definition. The quantities appearing in Ozawa's definition may be non-operational in the sense that they do not refer to observables which can be tested in the normal run of microscope measurements. But the linear operators involved have a clear meaning in the formalism and their expectations can obviously be measured in some way. That may be a challenging experiment of independent interest, but the failure of \eqref{false} is automatic if the implementation is correct. These experiments certainly do not help to prove Heisenberg wrong. They merely confirm that inequality \eqref{false}, which nobody ever claimed to be right, indeed isn't correct.

We hasten to add that, generally speaking,  we are very much in favour of foundational experiments. Only in this case a worthless target was chosen. It is true that sometimes a dubious theoretical calculation (say in condensed matter physics) may be given more credibility by confirming experiments. But in the present case such a correction by experiment simply does not happen. Ozawa's claims about Heisenberg are exactly as untenable after the experiments as they were before.

\subsection{How the debate might continue}
We certainly agree with Ozawa that the whole circle of ideas around uncertainty, and especially measurement uncertainty, deserves more rigorous elucidation. There will be many results contributing to this aim, and Ozawa's corrected relation may play a role there too. We also don't have a quarrel with state-dependent relations in this context, which may even be just the right thing for improving cryptographical security proofs. 
Naturally, the false relation \eqref{false} is simply to be ignored in this endeavour. We also do not expect that Ozawa's improved relation will play a very positive role, because it is based on ill-conceived basic notions. There are signs, however, that his group may be contributing after all. In a recent paper \cite{Buscemi} we see a lot of Ozawa's old rhetoric and misrepresentation of our work, but also a turn to an information-theoretic framework and state-independent quantities.

With better results appearing on the scene, criticising the noise operator approach may thus lose importance. So apart from Ozawa's direct attacks on our work (see below) there is only one point which should not go unchallenged: Ozawa's completely arbitrary attribution of the elementary mistake \eqref{false} to Heisenberg.

\section{Reply to Ozawa's criticism}

The aim of Ozawa's text \cite{Ozawa2013} seems not so much to criticize particular aspects of our paper \cite{BLW2013c} but to brand it as ``groundless'', as not covering some cases which ``escape it'' (despite claims to the contrary), as making ``unsupported claims'' and giving a ``failed proof''. It seems that Ozawa is not prepared to tolerate any countering in the literature to his claim of having refuted Heisenberg. Indicative of this is already the inversion of the title (his ``Disproving Heisenberg's error-disturbance relation'' vs.\ our ``Proof of Heisenberg's error disturbance relation''). It only shows that Ozawa is forgetting elementary logic in his rebuttal rage: Ozawa's paper is intended as a disproof only of our paper, so for his title to make any sense he would have to identify ``Heisenberg's error-disturbance relation'' with our approach, which is surely not what he wants to say.

Ozawa has submitted his paper to a journal whose editors asked us for an open comment to be transmitted to the referees and the author. This open comment is the basis of the current note, so we can assume that Ozawa has been informed of our arguments. Furthermore, in the meantime our paper with the full proof \cite{BLW2013b} has appeared on the arXiv (in December 2013). Neither text seemed sufficient reason for Ozawa to retract or even modify his arXiv submission \cite{Ozawa2013}.

\subsection{Ozawa on operationality}
Ozawa's paper does contain a reply to our criticism of his approach as non-operational (see above). This then does not refer to the paper \cite{BLW2013c}, which is the only source he cites: \cite{BLW2013c} focusses on establishing a positive result rather than criticising other approaches, and so contains nothing of the sort. But possibly Ozawa is referring here to meetings at conferences (RFW in Tokyo around 2002, PL at V\"axj\"o in 2013), and perhaps to the papers \cite{BuHeLa04,Werner04}, where we tried to explain this point to him. His resolution of the problem arising when the momentum before and after the measurement do not commute (p.1, top of first column) is that {\it sometimes they do}. Clearly, if that is the strongest answer he can give, his approach would have to relinquish all aspirations to ``universality''.  Again, this is hardly what he wants to say, but we have to put on record that our criticism remains unanswered.

\subsection{Linear models}
The particular example he cites here serves also as a ``counterexample'' to our theorem. Therefore we have to look at it more closely. Indeed, ``linear measurement models'' have served as a warm-up for the study of measurements since the time of von Neumann, and Ozawa has written several papers about them (cited in his manuscript). These models are relevant in the lab, for example in the quantum optics of field quadratures, where linear transformations are the easiest to perform, or in the mechanical case of a particle by scattering with another particle.

One can also describe this class abstractly  as canonical linear transformations of the joint phase space of particle and probe (Ozawa somewhat arbitrarily chooses a three-parameter subclass), which then directly translates into quantum mechanics, because all these transformations can be realized by quadratic Hamiltonians. Let us consider, for definiteness and for getting better assistance from  physical intuition, the mechanical case of a particle of mass $m_1$ whose position we determine by scattering of a test particle (``probe'') of mass $m_2$.

Ozawa considers a position measurement, whose outcome is directly the position of the probe. This is a linear function of the particle's initial position with slope $a=2m_1/(m_1+m_2)$. Now any lab assistant trying to calibrate this measurement would notice this factor, so in order to get a better measurement, one would naturally correct for it. Otherwise, the error on suitably chosen states will become arbitrarily large. In our notation, this translates to an infinite error (``worst case error'') of the position measurement. Including the correction, however, i.e., using an optimized estimator on the same data, one does get a measurement with typically finite error, depending on the initial state of the probe. For momentum one considers the particle's own momentum (not the probe's) after the measurement, which will be linear in the input momentum with slope $(m_1-m_2)/(m_1+m_2)$. For the disturbance to be finite we would need this factor to be 1, which contradicts $m_2>0$. Of course, this covers only the trivial part of the disturbance (momentum is different before and after) and not its ``uncontrollable'' aspect (momentum cannot be retrieved). A suitable correction can again be made and gives measurements with bounded momentum disturbance across all states.

Linear models can thus be used to produce good joint measurements of position and momentum, and as the dilation theory of covariant observables \cite{screen,QHA} shows, all phase space covariant measurements can be obtained in this way. These covariant measurements do play a key role in our theory, and indeed the measurement uncertainties are in this case directly related to the preparation uncertainties of the probe state, and with a minimal uncertainty initial probe state one also gets cases of saturation (equality) in our error-disturbance inequality.

Ozawa, however, takes not only these good measurements but also those in which the lab assistant failed to apply the appropriate correction factors. (Of course, as he notes in col2, p.3, the good ones are of Lebesgue measure zero in this set, which to him apparently suffices to brand them as an exotic specialty.) Anyhow, his claim that these models are reasonable and have finite errors, even with an ill chosen calibration factor, is entirely due to the difference between our state-independent and his state-dependent formulation and, in addition, to the ad hoc assumption of finite variances for the initial states. Perhaps we can follow him here by not calling the lab assistant infinitely stupid, but just say that his errors are unbounded. Of course, this in no way invalidates our result.

\subsection{The indeterminate case}
In the linear models one can easily produce examples where one uncertainty (in our sense) is zero and the other is infinite. In fact, it suffices to choose an ideal position measurement with a random momentum output, uncorrelated with the input state. Ozawa argues (in his criticism of Appleby, p.\ 3) that such cases per se invalidate any uncertainty relation, because the product $0\cdot\infty$ ``cannot be concluded to be above $\hbar/2$''. Here, as always in mathematics or physics, one has to discuss what indeterminate products mean. The point of the relation is that not both uncertainties can be simultaneously small. From this point of view the only reasonable demand concerning the indeterminate case would be that the theorem should ensure that if one factor is zero the other must be infinite. This is indeed covered by our result since if both uncertainties are finite or one of them zero the inequality holds. Then in whatever way one approaches a situation with one factor zero the other has to diverge, and not just arbitrarily, but in keeping with the finite-finite relation. Thus the indeterminate case requires no extra argument, provided that one does show the relation under the assumption that both factors are finite (including the case of one being zero).

It is perhaps interesting to consult Ozawa's handling of the indeterminate case in his own relations. Clearly, as the above examples show zero error or disturbance may occur. By his own logic any infinite terms, would invalidate his extended ``universally valid'' relation. So he just forbids ad hoc any states with infinite variance for position or momentum (at the bottom of p.~2 col.~1). To be sure, he vaguely refers to this as avoidance of ``inaccessible resources'', and we could counter his criticism in this spirit by saying that ``zero error is an unrealistic idealization'' anyhow. But with a proper understanding of the indefinite case no such rhetorical manoeuvers are needed.

\subsection{An alleged error in our argument}
The last column of Ozawa's paper is devoted to spotting an error in our argument. Indeed, since he feels that he has presented a counterexample, this is required if he wants to invalidate a proven theorem. Here we have to say that the proof in our paper \cite{BLW2013c} is clearly labelled as a sketch, in particular the averaging/compactness argument, which is described as somewhat technical. If anyone has doubts about a published proof (or claim thereof) there is a clear procedure dictated by the rules of scientific debate: Either one follows the hints given in the proof, including the hint that a very similar argument for just this part is to be found in an earlier paper \cite{Werner04}. Or one waits for the announced published full proof \cite{BLW2013b}. Or, failing the mathematical ability to do the first, and the patience to do the latter one can write to the authors requesting clarification. Ozawa has definitely not pursued one of these routes.

This would still be ok, if even from the sketch one could detect a fatal flaw in the argument. According to Ozawa the ``exact point where the argument fails'' is in our claim that the set of observables satisfying a certain calibration condition ``is convex and compact in a suitable weak topology''. He counters this merely by saying ``However, this is not true''. There is no argument whatsoever, beyond the remark that the closure of the set of {\it all} observables would contain ones supported on the Stone-\v Cech-compactification of $\Rl^2$ (actually it could be any other compactification depending on the weak topology which we did not even specify). He could well have learned this from the cited paper \cite{Werner04}, including a proof of how a subset as the one specified may be compact nevertheless. This is analogous to the space of density matrices, whose weak closure contains singular states such as states with sharp position. However, the subset on which the harmonic oscillator has fixed finite energy expectation is indeed weakly compact.

\subsection{An aspect where Ozawa could have made a point}
There is indeed a tension between our claim of an operational measure of errors and the fact that they may turn out infinite quite easily. Of course, this is due to taking the worst case over {\it all} quantum inputs over the measuring device. Thus in assessing the resolution of a microscope even objects in a neighbouring galaxy would be tested, which is clearly nonsensical. Ozawa does not make this point in the submitted paper. His claim that our quantities are so often infinite is based on the linear models (see above).

Anyhow, there is a non-operational aspect here in our work, which we intend to address in the near future.  What we are currently working to show is that the uncertainty relation as stated is valid even if we limit the calibration to some finite ``operating range''. Thus no states with arbitrarily large position or momentum would be involved in the definition of the errors. The relevant condition would be that these operating ranges in position and momentum are large compared to the respective errors. We expect that this would not substantially decrease the uncertainty bounds, with a difference going to zero in the limit of large operating ranges. The relation as stated would thus be interpreted as an idealized version where the operating ranges become infinite. In fact, Appleby \cite{Appleby1998b} has sketched a similar construction in the noise operator approach.

Even in this more realistic version, however, the distinction made in the first section would persist: It makes no sense to limit the ``operating range'' to a single state if one aims to define error measures as figures of merit for a measuring device.

\section{Reply to Rozema {\em et al} \cite{Roz13}}

This section comments briefly on the submission \cite{Roz13} of  Rozema {\it et al} entitled
``A Note on Different Definitions of Momentum Disturbance''. While also apparently conceived as a rebuttal of our paper \cite{BLW2013c}, its tone is rather more conciliatory. We are certainly sympathetic to the comparative approach indicated in the title. However, there are several misunderstandings that need clarification.
Our comment was also originally drafted following a request by an editor to comment on the submission \cite{Roz13} so that we may expect that the authors are informed on our critique.
However, the authors have remained quiet, no revision of \cite{Roz13} has appeared.

\subsection{Once again: error as figure of merit}

The motivation for building figures of merit for measuring devices (see the introduction) was apparently not appreciated by the authors of  \cite{Roz13}. We consider quantities like the resolution of a microscope, which does not depend on the object we are looking at: It is a promise for accuracy on arbitrary input states. What remains of this in the presentation of \cite{Roz13} is that we consider the ``disturbing power'' of an instrument, while the disturbance on a particular state may be smaller. We agree, but that is besides the point for a benchmark quantity. In the same spirit one could propose not to talk about the resolution of a microscope but about its ``error generating capacity'', which is presumably an irrelevant quantity because it does not refer just to the one object at hand but is attained at a quite different one.

In this light, the conclusion shortly after equation (2) of \cite{Roz13}, namely that ``their relationship does not hold
for a given state'', may suggest itself because no state appears in it. Logically, however, since our quantities are maximized over all states, our relation does in fact hold for arbitrary states -- it simply does not give the tightest estimate of errors specific to any particular state. Moreover, it does make a statement about every possible joint measurement device, and hence applies to every experiment involving such devices. Of course, if one is interested in measurement error and disturbance measures for a single state, one would turn to state-dependent measures instead of our apparatus benchmark quantities; in fact, our non-maximized quantities are operationally significant as measures of error and disturbance for an individual state -- in contrast to Ozawa's measures.

Maybe because the authors have not fully appreciated the basic nature of our quantities, they misrepresent our result as the conclusion that
``any measurement which is capable of achieving a measurement precision of $\Delta X$ on {\em some} states must be capable of imparting a momentum disturbance of $\hbar/2\Delta X$ on {\em some} (potentially different) state''.  (Even though this is presented as a quote, it is not a quote from our paper).
The emphasis of the first {\em some}  is ours: this is plain wrong, it should rather read {\em ``applicable to all''}. The second {\em some} is correct, but the phrase ``is capable of'' is a very odd rendering of the correct ``necessarily imparts''. The parenthetical ``potentially different'' should be deleted, because it makes no sense when the premise is for {\em all} states.

\subsection{Not all measurements prepare (approximate) eigenstates}

In the last paragraph of description of our work (p.~1, col.~2) Rozema {\em et al} convey the impression that our work is mathematically trivial, by saying that we would just need to realize that after a position measurement of accuracy $\Delta X$, the post measurement state would necessarily have a position spread $\Delta X$ (and hence have the claimed momentum uncertainty). But this assumption about the post-measurement state is entirely unwarranted, and we certainly do not make it.

Heisenberg in his 1927 paper has clearly made this identification of position measurement error and the position uncertainty in the final state, in line with the common but limited perspective at the time, that measurements always produce (approximate) eigenstates. It has long since become clear that this identification of the concept of measurement with projective, or von Neumann-L\"uders measurements is too narrow. Indeed, for such measurements the preparation uncertainty relation will trivially entail the joint measurement error relation. Interestingly, it turns out that this logical connection does persist in the case of general joint measurements, as we have shown in a variety of examples in \cite{BLW2013b,BLW2014}.

\subsection{Comparing Ozawa's ``error'' with actual statistical deviations}

It is quite disconcerting that there has not been any critical conceptual analysis of Ozawa's measures of disturbance and error
among those research groups that use them. Rozema {\em et al} share with Ozawa the belief that ``a good measure of the disturbance to a state is the root-mean-squared (RMS) difference between $P$ before and after the process, $U$:
$$
\eta_O(P)=\langle(U^*PU -P)^2\rangle^{\frac 12}.\qquad\qquad (3)\text{''}
$$
If Quantum Mechanics were Classical Mechanics this would be fine. But Quantum Mechanics is not Classical Mechanics.  In Quantum Mechanics the operationally significant measurement result is not an individual outcome but a distribution of outcomes obtained when the same measurement is performed many times on the same system prepared in the same way (or, if one prefers, when the measurement is performed in a similarly prepared ensemble of (independent) systems). Therefore, to compare the momentum before and after the process, {\em one has to compare the two momentum distributions, the momentum distribution before the position measurement and the momentum distribution after the position measurement}.

To this end we use the Wasserstein distance of order 2  \cite{Villani}, which does constitute an operationally significant extension of the classical RMS difference to the noncommutative context.
Only in some special cases is Ozawa's measure $\eta_O(P)$  operationally accessible from the data available (which are the statistics of the approximate measurement at hand and that of an accurate reference measurement performed on an ensemble prepared in the same state), and even in those cases it usually overestimates the disturbance in the momentum distribution (since in the case of the order 2 Wasserstein distance one minimizes the mean of the squared deviation over all joint distributions while for Ozawa's measure one just picks a particular joint distribution).

This is made evident in the example presented by Rozema {\em et al}: the ``difference'' between the position observable and the space-inverted position observable can be noticed in some states but not in  states with a parity symmetric distribution of the position. This means simply that the state change due to a parity operation simply cannot be detected by the position observable in all possible states, and accordingly our measure of the distance between the two (identical) distributions to be compared vanishes appropriately in the symmetric case while Ozawa's measure gives a misleading nonzero value.

We have expressed our dissatisfaction with Ozawa's definition only in passing in our paper \cite{BLW2013c} pointing to a more detailed discussion in preparation, which is now available \cite{BLW2013a}.
Rozema {\em et al} cite a 2004 paper \cite{Werner04} 
in this regard, which contains  the statement that Ozawa's definition is not operational. It appears that Rozema {\em et al} consider this refuted by having measured  Ozawa's quantities. That is, of course, not correct. In principle, any operator, like $ U^*PU - P$,
can be measured, be it by realizing an instrument, which determines the outcomes for all spectral projections, or, in the case at hand, in a less detailed way by a weak measurement. The charge is that a meta-experiment is required, and that disturbance and error are not meaningful in terms of the device as given.  As noted
above, Ozawa has implicitly conceded this point in his ``disproving'' paper, where he counters this objection only by giving an example  of a special case when it does not apply.
The criticism, expressed by way of examples also as early as 2004 in \cite{BuHeLa04},
that Ozawa's error (and similarly then disturbance) can even vanish in states where the input and output distributions are clearly vastly different has never been refuted or directly commented upon up to now. Further striking examples of this kind, including the analogous situation of Ozawa's disturbance vanishing where the observable in question does get disturbed, have been given in \cite{BLW2013a}.

Returning briefly to the quantity $\eta_O(P)$, the definition recalled above makes it evident that, unless the momentum and the disturbed momentum commute, the quantity $U^* PU-P$ does not commute with either of the two momentum observables, and therefore its measurement is not compatible with the measurement of the undisturbed or disturbed momentum. Hence there is no reason to expect that its values signify the deviation between the undisturbed and disturbed momentum, not any more than measuring kinetic and potential energy separately constitute a measurement of the Hamiltonian of a quantum particle. A similar remark applies to Ozawa's error.

It is surprising that the experimental groups that succeeded in carefully measuring $\eps_O(A)$ and $\eta_O(B)$ as quantum mechanical expectation values made no attempt to confront these expectation values with the actual error  in an approximate measurement of $A$ and the associated actual disturbance of $B$ in the individual state. The interpretation of Ozawa's quantities is always simply taken for granted -- which is understandable to a degree due to their intuitive formal appeal. But it remains puzzling that their deficiencies go unnoticed in the discussion of the experiments although they show up unmistakably (see below).

\subsection{On trivially vanishing error products (again)}

In the same paper \cite{Werner04} Ozawa's definition was actually given the benefit of  doubt. It is mathematically correct that a state-dependent relation could be stronger by giving more detailed information, from which our ``worst case'' statement would trivially follow by taking a maximum. The catch here is (back then as now) that the state-dependent analogs are just not true, as Ozawa himself showed, and as was (entirely unsurprisingly) verified by the recent experiments.  Indeed the failure of Heisenberg type relations in the Ozawa approach is a trivial consequence of the state dependent definition, and can be realized by models as simple as an exact position measurement, followed by an exact repreparation of the input state. There is nothing physically interesting to be understood from this failure, even if the experiments as such may have been demanding. This ``failure'' also applies to our operationally significant measures, as witnessed by the model example just referred to \cite{BLW2013a}. This said, we rush to add  that a study of state-dependent error measures, like the Wasserstein-2 distance, is needed, for instance, if one wants to know how much error one has to allow for a measurement in a particular state if one wants to impose a specific bound to the disturbance of some observable in that state.

\subsection{Ozawa's inequality destroys its own (intended) interpretation}

That the interpretation of $\eta_O(P)$ as a measure of disturbance is ultimately untenable can be seen most strikingly, we believe, in the following example. The quantum theory of measurement asserts the existence of error-free measurement of position with the property that the associated nonselective channel is a constant map, $\rho\mapsto\rho_0$. In this case
Ozawa's inequality [eq. (5) in \cite{Roz13}] reduces to
\[
\eta_O(P)\ \ge\ \frac\hbar{2\Delta X}.
\]
Now, consider the situation where the input state $\rho$ is chosen to be identical to the fixed output state,
$\rho=\rho_0$. In that case there is virtually no disturbance for this state, and yet the above inequality shows that $\eta_O(P)$ can be as large as one wishes (by choosing $\rho_0$ with sufficiently small $\Delta X$. By contrast, again our measure of disturbance vanishes in this case.
Hence one might say that Ozawa's inequality invalidates its own (intended) interpretation.

We are pleased to see that, in contrast to Ozawa, Rozema {\em et al} do not claim to have read Heisenberg's mind. Yet they conclude with the statement that although Ozawa's relation contradicts Heisenberg's, ``we believe that it is closer in spirit to the disturbance {\em typically associated} with Heisenberg's microscope than the definition of Busch et al.''  (Our emphasis). We are not so sure about the silent majority, but we trust that the matter can be clarified by scientific debate.

In fact, a curious convergence of the state-independent and state-dependent perspectives on measurement errors emerges in the qubit case. This becomes apparent by a closer look at the  Vienna and Toronto experiments, which also serves to illustrate some of the shortcomings of Ozawa's ``error'' highlighted above.

\subsection{What the Vienna and Toronto experiments actually show}

It must indeed be noted that the experiments of the Vienna \cite{Erh12} and Toronto \cite{Roz12} groups are not about the Heisenberg uncertainty relations for canonical conjugate variables, but about certain qubit measurements. The Ozawa relations can be stated in this context, but so can our approach. In fact, it becomes technically much easier. For example, the compactness argument Ozawa tried to take issue with is not necessary at all. To facilitate the comparison we have in the meantime spelled out the finite dimensional version of our approach, particularly for qubit systems \cite{BLW2014}. We have obtained a joint measurement error trade-off relation of the Heisenberg type, which is of the following generic form: if observables $A,B$ are to be approximated in terms of jointly measurable observables $C,D$, respectively, then the errors $\Delta(A,C)$ and $\Delta(B,D)$ obey an inequality
\begin{equation}\label{qubit-ur}
\Delta(A,C)^2+\Delta(B,D)^2\ge (\text{\rm incompatibility of $A,B$}).
\end{equation}

Now it turns out that for qubit observables of the form investigated in the experiments, Ozawa's quantities
$\epsilon_O$ and $\eta_O$ do not depend on the input state; this undermines their interpretation as state-specific error or disturbance measures.  In both experiments, $\epsilon_O$ and
$\eta_O$ are generally bad over-estimates of the actual state-dependent errors. Curiously, the authors of \cite{Vienna2} do notice the state-independence of $\epsilon_O$, and they do notice that in some states the measured distribution is identical to that of the target observable, even where these observables do not commute; but they do not seem to consider this as an indication of the failure of $\epsilon_O$ as a
state-specific error measure.

The quantity $\epsilon_O$ turns out to be directly related to our state-independent error measure and can therefore serve as a rough estimate for it in the qubit case.
As a consequence of  this fact and \eqref{qubit-ur}, we found  further that Ozawa's quantities obey a similar trade-off relation if applied to the quantities $A,B$ and their jointly measurable approximators $C,D$:
\begin{equation}\label{qubit-oz-ur}
\epsilon_O(A,C)+\epsilon_O(B,D)\ge \tfrac12(\text{\rm incompatibility of $A,B$}).
\end{equation}
Thus, irrespective of the fact that these quantities fail as faithful state-dependent error measures, they nevertheless give rise to a Heisenberg-type trade-off in the qubit case! This clearly shows that Ozawa's inequality, rather than constituting a violation of Heisenberg's error-disturbance relation, simply fails to capture its spirit. The root of this failure can be seen in the unsuitability of the error product to describe such trade-off relations in the case of discrete observables.

We have also shown that the Toronto experiment, which was conceived as a weak measurement, can be
interpreted as an approximate joint measurement of smeared versions $C,D$ of qubit observable $A,B$.
The values of $\epsilon_O(A,C)$ and $\eta_O(B,D)\equiv\epsilon_O(B,D)$ are there actually equal to our errors,
$\Delta(A,C)$ and $\Delta(B,D)$; hence for this measurement scheme, Ozawa's quantities also satisfy the inequality \eqref{qubit-ur}. Interestingly, for certain parameter settings the Toronto experiment
is found to saturate the bound in this inequality. In fact, we would not be surprised if the data for
$\epsilon_O(A,C)$ and $\epsilon_O(B,D)$ obtained in the experiments were already sufficient to check this inequality.

\bibliographystyle{unsrt}

\end{document}